\documentclass{emulateapj}
\usepackage{graphicx}
\usepackage{hyperref}
\begin{document}
\slugcomment{To be published in The Astrophysical Journal Letters}
\title{First asteroseismic limits on the nature of dark matter}
\author{Jordi Casanellas\altaffilmark{1,3}, Il\'\i dio Lopes\altaffilmark{1,2,3}}
\altaffiltext{1}{CENTRA, Instituto Superior T\'ecnico, Lisboa, Portugal} \altaffiltext{2}{Departamento de F\'isica, Universidade de \'Evora, Portugal} 
\altaffiltext{3}{E-mails: jordicasanellas@ist.utl.pt, ilidio.lopes@ist.utl.pt} 
\shorttitle{First asteroseismic limits on the nature of dark matter}
\shortauthors{Casanellas and Lopes}
\begin{abstract} 
We report the first constraints on the properties of weakly interacting low-mass dark matter (DM) particles using asteroseismology. The additional energy transport mechanism due to accumulated asymmetric DM particles modifies the central temperature and density of low-mass stars and suppresses the convective core expected in 1.1-1.3 M$_{\odot}$ stars even for an environmental DM density as low as the expected in the solar neighborhood. An asteroseismic modeling of the stars KIC 8006161, HD 52265 and $\alpha$ Cen B revealed small frequency separations significantly deviated from the observations, leading to the exclusion of a region of the DM parameter space mass versus spin-dependent DM-proton scattering cross section comparable with present experimental constraints.
\end{abstract}
\section{Introduction}
\label{sec-intro}
The identification of the nature of the dark matter (DM) of the Universe is a major open problem in modern physics \citep{Bertone:2010at}. Among the diverse strategies for DM searches, the study of the possible impact of DM in the properties of stars has been explored in recent years as a complementary approach to the DM problem~\citep{Spolyar:2007qv,Scott:2008ns,Casanellas:2009dp,Zackrisson:2010jd,Sivertsson:2010zm,Casanellas:2011qh,Scott:2011ni,Li:2012qf,2012MNRAS.422.2164I,Corsico:2012ki}. In particular, weakly interacting DM candidates with an intrinsic matter-antimatter asymmetry~\citep{Kaplan:2009ag,Davoudiasl:2011fj,Blennow:2012de} do not annihilate after gravitational capture by compact astrophysical objects and can therefore strongly influence their internal structure 
\citep{1987NuPhB.283..681G}. 
Thus, both the observation or the lack of observation of the impact of asymmetric DM (ADM) on the properties of stars can be used to put constraints on the characteristics of these DM candidates.

The interior of the Sun, being known with a high accuracy thanks to solar neutrinos and helioseismic data, is an excellent laboratory to probe the existence and the properties of ADM particles. Such particles remove energy from the inner $\sim$4\% of the Sun, leading to a reduction of the central temperature and the creation of an isothermal core~\citep{art-Taosoetal2010PhRvD,art-FrandsenSarkar2010PhRvL,Lopes:2012af}. In particular, ADM candidates with low-masses and large spin-dependent (SD) proton scattering cross sections may influence the internal solar structure so strongly that they would produce clear signatures in the low-degree frequency spacings and in the solar gravity 
modes~\citep{Lopes:2010fx,Cumberbatch:2010hh,2012ApJ...746L..12T}. Interestingly, low-mass weakly interacting massive particles (WIMPs) 
with similar characteristics provide an explanation for the signals in various direct detection experiments, strengthening the motivation for the search of indirect signatures of these particles.

It has also been shown that these low-mass ADM candidates may produce marked effects in very low-mass stars and brown dwarfs~\citep{Zentner:2011wx}. In environments with high ADM densities, solar-like stars may show significant deviations in their evolutionary tracks~\citep{Iocco:2012wk}. Also neutron stars, due to their compactness, capture DM very efficiently and may be strongly influenced by the accumulation of ADM~\citep{Bertone:2007ae,Kouvaris:2011fi,art-LeungChuLin2012}. Here we will show that, even for a DM density as low as the expected in the solar neighborhood, $\rho_\chi=0.4\;$GeV cm$^{-3}$~\citep{Garbari:2012ff}, main-sequence stars with masses similar to that of the Sun present distinct signatures of the captured ADM.

With the advent of asteroseismology, a precious insight into the stellar interiors is nowadays possible for the first time. The \textit{CoRoT}~\citep{Michel:2008im} and \textit{Kepler}~\citep{Bruntt:2012hs} missions have already detected oscillations in about 500 stars~\citep{Chaplin:2011wa}. This fact has allowed to test theories of stellar evolution and to probe the stellar cores with an unprecedented precision~\citep{Garcia:2010rn,Bonaca:2012av}. The seismic analysis of stars other than the Sun is complementary to helioseismic DM searches because it allows the study of stars with lower masses, which are more strongly influenced by DM, and stars whose dominant energy transport mechanisms may change due to the DM influence. In this Letter we will demonstrate, by studying the case of the stars KIC 8006161, HD 52265 and $\alpha$ Cen B, that present asteroseismic observations do constrain a significant region of the DM parameter space.

\section{Interaction of dark matter and stars}
\label{sec-DMstars}
\begin{deluxetable*}{l r r r r r r r}
\tablewidth{0pt}
\tablecaption{Constraints on the stellar characteristics adopted for the modeling and selected results.\label{tab-starchars}}
\tablehead{Star & $M$ (M$_{\odot}$) & $R$ (R$_{\odot}$) & $L$ (L$_{\odot}$) & $T_{eff}$ (K) & $(Z/X)_s$ & $\langle\Delta \nu_{n,0}\rangle$\tablenotemark{a} ($\mu$Hz) & $\langle\delta \nu_{02}\rangle$\tablenotemark{a} ($\mathit{\mu}$Hz)}
\startdata
\multicolumn{2}{l}{\textbf{KIC 8006161}} &&&&&& \\
\hspace{0.1cm}Observations\tablenotemark{b} & 0.92-1.10 & 0.90-0.97 & $0.61 \pm 0.02$ & $5340 \pm 70$ & $0.043 \pm 0.007$ & $148.94 \pm 0.13$ & $10.10 \pm 0.16$ \\
\hspace{0.1cm}Stand. modeling & 0.92 & 0.92 & 0.63 & 5379 & 0.039 & 149.03  & 10.12 \\
\hspace{0.1cm}DM modeling\tablenotemark{c} & 0.92 & 0.92 & 0.63 & 5379 & 0.039 & 149.08  & 9.13 \\
\multicolumn{2}{l}{\textbf{HD 52265}} &&&&&& \\
\hspace{0.1cm}Observations\tablenotemark{b} & 1.18-1.25 & 1.19-1.30 & $2.09 \pm 0.24$ & $6100 \pm 60$ & $0.028 \pm 0.003$ & $98.07 \pm 0.19$ & $8.18 \pm 0.28$ \\
\hspace{0.1cm}Stand. modeling & 1.18 & 1.30 & 2.22 & 6170 & 0.028 & 97.92 & 8.16 \\
\hspace{0.1cm}DM modeling\tablenotemark{c} & 1.18 & 1.30 & 2.22 & 6170 & 0.028 & 98.05 & 7.65 \\
\multicolumn{2}{l}{\textbf{$\alpha$ Cen B}} &&&&&& \\
\hspace{0.1cm}Observations\tablenotemark{b} & $0.934 \pm 0.006$ & $0.863 \pm 0.005$ & $0.50 \pm 0.02$ & $5260 \pm 50$ & $0.032 \pm 0.002$ & $161.85 \pm 0.74$ & $10.94 \pm 0.84$ \\
\hspace{0.1cm}Stand. modeling & 0.934 & 0.868 & 0.51 & 5245 & 0.031 & 162.56 & 10.23 \\
\hspace{0.1cm}DM modeling\tablenotemark{c} & 0.934 & 0.868 & 0.51 & 5230 & 0.031 & 162.45 & 8.95 \\
\enddata
\tablenotetext{a}{Averages for the intervals $2750<\nu$($\mu$Hz)$<3900$ (KIC 8006161), $1600<\nu$($\mu$Hz)$<2600$ (HD 52265), and $3300<\nu$($\mu$Hz)$<5500$ ($\alpha$ Cen B).}
\tablenotetext{b}{Data from \cite{Mathur:2012sk} and \cite{Bruntt:2012hs} (for KIC 8006161), \cite{Ballot:2011kn} (for HD 52265), and \cite{Kjeldsen:2005td} (for $\alpha$ Cen B).}
\tablenotetext{c}{$m_{\chi}=5\;$GeV, $\sigma_{\chi,SD}=3\times10^{-36}\;$cm$^2$, $\rho_{\chi} = 0.4\;$GeV cm$^{-3}$.}
  \end{deluxetable*}
Nearby stars are embedded within the halo of DM particles that is presently believed to permeate our Galaxy. If these DM particles have a non-negligible scattering cross section off baryons (so they are WIMPs), then some of them may collide with the nucleons of the stellar plasma, losing kinetic energy. A fraction of these DM particles is gravitationally captured by the stars. To calculate the capture rate we follow the formalism that was early developed by \cite{art-Gould1987}, as implemented in \cite{art-GondoloEdsjoDarkSusy2004}. We assume a Maxwell-Boltzmann distribution of the velocities of the DM particles, 
 with a dispersion $\bar{v_{\chi}}=270\;$km s$^{-1}$, and a stellar velocity of $v_{\star}=220\;$km s$^{-1}$. The expected deviation from the mentioned fiducial values for the specific stars studied in this work may lead to a maximum error on the capture rate of approximately 15\% 
 (see \cite{Lopes:2011rx} for details). 
 
In the asymmetric WIMP scenario, the annihilation cross sections required to match the DM relic density are larger than in the standard WIMP scenario, depending on the degree of asymmetry \citep{Iminniyaz:2011yp}. Nevertheless, in the asymmetric scenario the particles concentrated on the stellar core cannot find a partner with whom to annihilate (the DM particle is not its own anti-particle) and therefore their number grows indefinitely while more particles are being captured. The same efficiency in the stellar accumulation of DM occurs for very feebly annihilating ($\langle \sigma_a v \rangle \lesssim 10^{-33}\;$cm$^3\;$s$^{-1}$) Majorana DM particles. In other cases, the captured DM would have no relevant impact on nearby low-mass stars.
 
The evaporation of DM particles can be neglected for the ADM candidates and stars considered in this work: while a low stellar mass tends to favor evaporation, this fact is compensated by the cooler stellar temperatures, $Evap\propto e^{-G M m_{\chi} / R T}$~\citep{1987NuPhB.283..681G}, in agreement with the results of \cite{Zentner:2011wx} for $\sim0.1\;\textrm{M}_{\odot}$ stars.

The DM particles captured in the stellar core provide a new energy transport mechanism that removes energy from the center of the star. The efficiency of this mechanism depends mainly on the ratio between the mean free path of the WIMPs inside the stellar plasma $l_{\chi}$ and the characteristic radius of the WIMPs distribution in the core of the star $r_{\chi}$~\citep{1986ApJ...306..703G}. For most of the WIMP-proton SD scattering cross sections $\sigma_{\chi,SD}$ considered here, $l_{\chi}>r_{\chi}$ and the energy transport by WIMPs is non-local. On the other hand, for large values of $\sigma_{\chi,SD}$, $l_{\chi}<r_{\chi}$ so the WIMPs are in local thermal equilibrium with the baryons. 
The latter regime applies only to values of $\sigma_{\chi,SD}$ which are not considered in this work ($\sigma_{\chi,SD}\gtrsim10^{-33}\;$cm$^2$). However, we follow the prescription described in \cite{Gould:1989ez} that extends the formalism developed for the local thermal equilibrium to other regimes by the use of tabulated suppression factors.
 
The DM capture and energy transport mechanisms were implemented in \texttt{CESAM}~\citep{1997A&AS..124..597M}, a sophisticated stellar evolution code. In the case of the Sun, the results of our modified solar model (\textit{e.g.} \cite{Lopes:2012af}) are in agreement with those of other codes in the literature~\citep{art-Taosoetal2010PhRvD,art-FrandsenSarkar2010PhRvL}. The observational constraints used for the modeling of the stars KIC 8006161, HD 52265, and $\alpha$ Cen B, as well as the results of some selected models with and without taking into account the DM effects, are summarized in Table~\ref{tab-starchars}.
  
\section{Impact of ADM on the properties of low-mass stars}
\label{sec-impact}
\subsection{Modifications of central temperature and density}
\begin{figure}[!t]
\centering
 \includegraphics{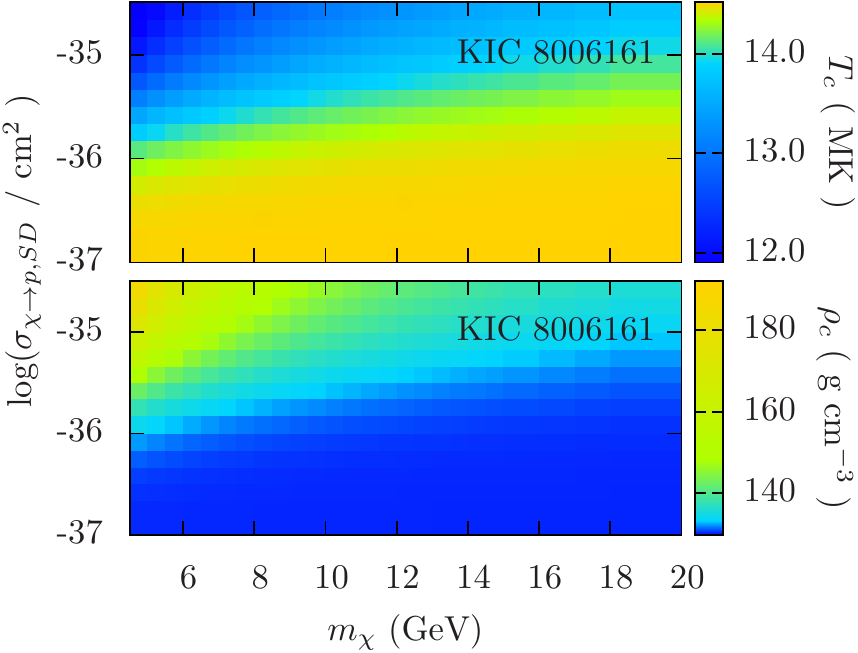}
 \caption{Central temperatures (top) and densities (bottom) of the DM-modified stellar models that reproduce the observed properties of the star KIC 8006161.\label{fig-Tc}}
 \end{figure}

The main signature of the additional DM cooling mechanism is a decrease in the central temperature and an increase in the central density. These variations are shown in Figure~\ref{fig-Tc} for several DM-modified stellar models, calibrated to reproduce the observed properties of the star KIC 8006161, for a range of DM masses and SD scattering cross sections. Compared with the standard modeling, for $m_{\chi}=5\;$GeV and $\sigma_{\chi,SD}=3\times10^{-36}\;$cm$^2$ we found a $\sim9\%$ decline in the central temperature. The variations on the internal properties are larger than those reported in the case of the Sun~\citep{art-Taosoetal2010PhRvD} because the importance of the energy transported by the WIMPs ($\varepsilon_{\chi,trans}\propto C_{\chi} \propto M_{\star}$) over the thermonuclear 
energy ($\varepsilon_{nucl} \propto M_{\star}^{3.5}$) increases when the stellar mass decreases. In particular, in our computations we found the DM cooling to reduce the $T_c$ of 0.7 M$_{\odot}$ stars nine times more efficiently than for 1.1 M$_{\odot}$ stars. This fact reinforces the potential advantages of performing DM searches in stars other than the Sun.
 
\subsection{Suppression of convective core}
In the standard picture of stellar evolution, stars with masses greater than 1.1 M$_{\odot}$ are expected to keep a convective core during most of the main sequence, while stars with lower masses quickly lose their convective cores. Convection arises when the gradient of temperature in the core is so steep that a rising bubble of plasma does not cool enough with its adiabatic expansion, so that it continues to rise, leading to a convective instability. If the temperature gradient is reduced by an additional mechanism such as the energy transport by WIMPs, then the conditions for convection may no longer be achieved. This possibility was first suggested in \cite{1987A&A...171..121R}, where the suppression of convection in horizontal branch stars was predicted using analytical approximations. This scenario must not be confused with the creation of an unexpected convective core in 1 M$_{\odot}$ stars due to the self-annihilation of DM particles captured in 
halos with very high DM densities~\citep{Casanellas:2010he}.

\begin{figure}[!t]
\centering
\includegraphics{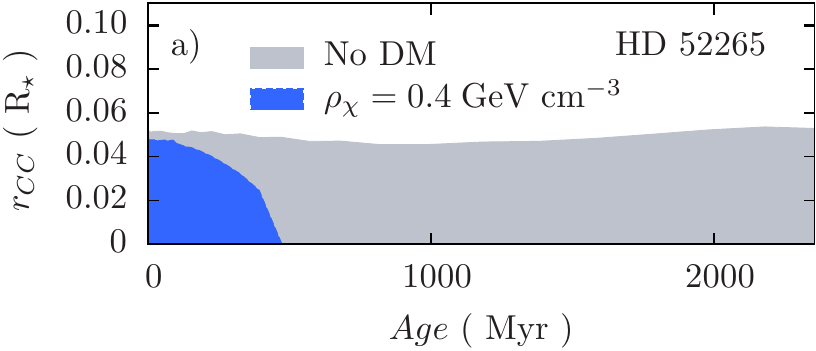}
\includegraphics{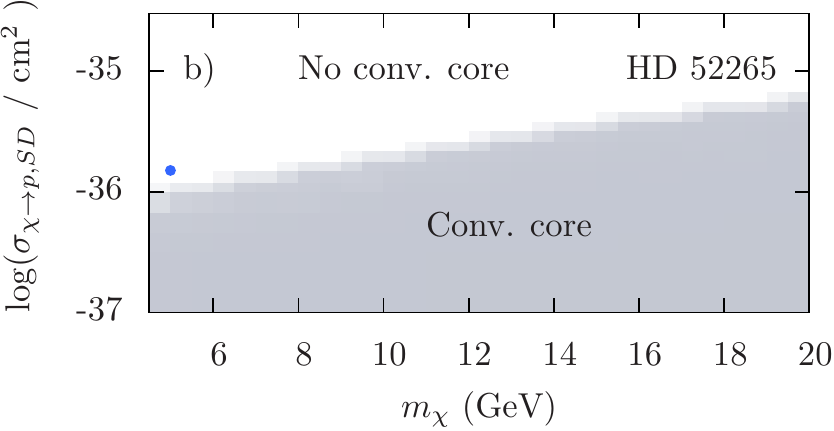}
\caption{(a) Size and duration of the convective core in the modeling of the star HD 52265 in the classical picture (gray) and taking into account the energy transport due to ADM particles with $m_{\chi}=5\;$GeV and $\sigma_{\chi,SD}=1.5\times10^{-36}\;$cm$^2$ (blue). (b) The presence of a convective core in HD 52265 depends on the mass and SD scattering cross section of the DM particles.\label{fig-CC}}
 \end{figure}

The reduction of the temperature gradient in the stellar interior due to the additional cooling by WIMPs was found to suppress the convective core expected in stars with masses slightly greater than that of the Sun. The standard modeling of the star HD 52265 predicted a convective core during all the main sequence, but this convective core rapidly disappeared when the energy transport by WIMPs was taken into account (see Figure~\ref{fig-CC}(a)). The range of DM masses and SD scattering cross sections for which the suppression of the HD 52265 convective core is expected is shown in Figure~\ref{fig-CC}(b)).  Interestingly, hints of the signatures of a convective core in HD 52265 were reported in \cite{Ballot:2011kn}. However, no conclusive information can be extracted until there is no definitive diagnostic of its presence or its absence (see also \cite{Escobar:2012qb}).
 
\section{Asteroseismic diagnostic of the presence of DM}
\label{sec-astroseism}
The characteristic signatures reported in the last section are potentially detectable with the analysis of the stellar oscillations. Asteroseismology is presently showing its power in determining with high precision not only the global properties of stars but also their internal structure. In particular, the small frequency separations of low angular degree ($l=0$) and radial order $n$, $\delta \nu_{02}=\nu_{n,0}-\nu_{n-1,2}$, have been shown to provide useful information about the core of the stars~\citep{1986hmps.conf..117G}. Thus, we would expect the seismic parameter $\langle\delta \nu_{02}\rangle$ to be sensitive to the modifications introduced by the WIMPs on stars.

We have computed the oscillation frequencies and separations of the DM-modified stellar models of KIC 8006161, HD 52265 and $\alpha$ Cen B using the \texttt{ADIPLS} package~\citep{art-Ch-Dals2008Ap&SS}. In order to disentangle the effects of DM from those arising from the variation of the stellar parameters, a very precise determination of the latter is of utmost importance. Although asteroseismology has already provided very accurate measurements of the mass and radius of KIC 8006161~\citep{Mathur:2012sk} and HD 52265~\citep{Escobar:2012qb}, with uncertainties of the order of 1\%, we preferred to focus here on the case of $\alpha$ Cen B, a star whose fundamental parameters are independently measured with high precision (see Table~\ref{tab-starchars}) thanks to its proximity and because it belongs to a binary system. The mass of this star has been determined from the radial velocities of $\alpha$ Cen A and B, its effective temperature from high quality spectra, its luminosity from photometric data and its 
radius from measurements of its angular 
diameter combined with parallax (see \cite{Kjeldsen:2005td} and references therein).

\begin{figure}[!t]
 \centering
 \includegraphics{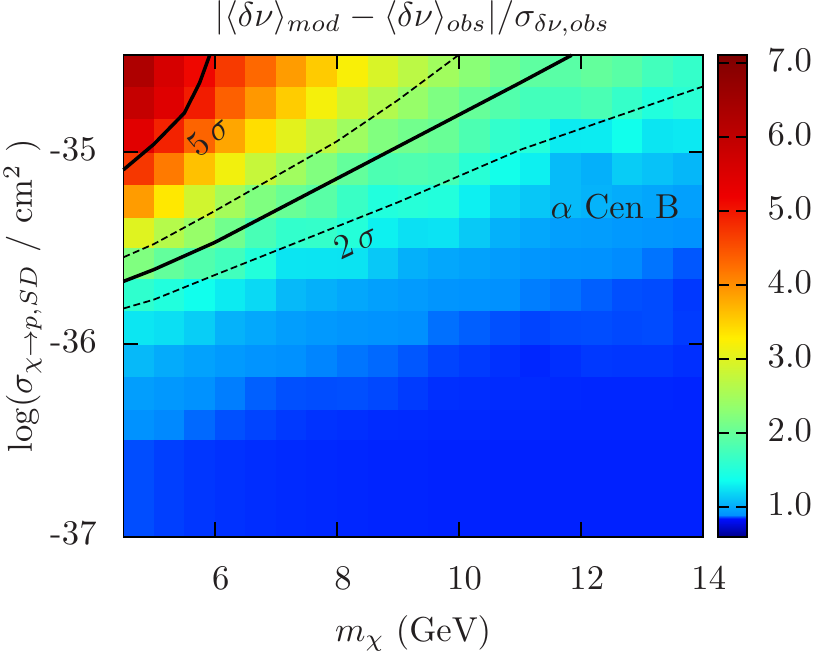}
 \caption{Deviation of the small frequency separation $\langle\delta \nu_{02}\rangle$ of the DM-modified stellar models from the true value measured in $\alpha$ Cen B. All the stellar models are calibrated to fit the $M$, $R$, $L$, $T_{eff}$, $(Z/X)_s$ and $\langle\Delta \nu_{n,l}\rangle$ of $\alpha$ Cen B within the observational error. The dashed black lines around the $2\sigma$ line show the uncertainty in $\langle\delta \nu_{02}\rangle$ arising from the observational error in the stellar characteristics.\label{fig-ssACB}}
\end{figure}

All stellar models used to create Figure~\ref{fig-ssACB} reproduce the measured $M$, $L$, $R$, $T_{eff}$, $(Z/X)_s$ and mean large frequency separation $\langle\Delta \nu_{n,l}\rangle$ of $\alpha$ Cen B within the observational error. However, while models without DM are also able to reproduce the observed mean small frequency separation $\langle\delta \nu_{02}\rangle$, we found that the stellar models with a strong influence of DM predict a $\langle\delta \nu_{02}\rangle$ significantly deviated from the observed value. The black lines in Figure~\ref{fig-ssACB}, labeled $2\sigma$ and $5\sigma$, show the DM characteristics corresponding to the calibrated models that predicted a $\langle\delta \nu_{02}\rangle$ with a difference of two and five times the observational error, respectively, from the observed value. 
The dashed black lines around the $2\sigma$ line show the uncertainty in the modeling when the observational errors in the stellar characteristics $M$, $L$, $R$, $T_{eff}$ and $(Z/X)_s$ are taken into account. 
This uncertainty corresponds to the standard deviation on $\langle\delta \nu_{02}\rangle$, evaluated computing 2600 valid models of $\alpha$ Cen B and including also the uncertainty in the capture rate $C_{\chi}$ from variations in the DM halo parameters and the stellar velocity, as discussed in Section~\ref{sec-DMstars}. The dashed lines around the $5\sigma$ line are not shown for clarity, but they would appear narrower because $\langle\delta \nu_{02}\rangle$ varies more abruptly in that region of the plot. Therefore, we conclude that present asteroseismic measurements of $\alpha$ Cen B disfavor the existence of DM particles with parameters above the $2\sigma$ line with 95\% confidence level.

Similarly, the presence of a convective core leads to strong asteroseismic signatures. The mixing of elements in convective regions introduces sharp structural variations in the border with radiative regions that produce a clear oscillatory signal in the frequency spectrum. It has been shown that this feature may be used to detect and measure the size of a convective core through asteroseismic parameters such as $r_{01}$, $r_{10}$ or $dr_{0213}$~\citep{2011A&A...529A..10C,2011A&A...529A..63S}. If these asteroseismic
diagnostic tools succeed in the confirmation of the presence or the absence of a convective core in a star with 1.1-1.3 M$_{\odot}$, this hypothetical measurement may be used to place further constraints on the nature of the DM particles. The characteristic and localised effects of DM should allow the disentanglement of its signatures from standard processes. Remarkably, several stars with the appropriate characteristics are presently being observed by the \textit{CoRoT} and \textit{Kepler} missions.

\section{Conclusions and Discussion}
\label{sec-concl}
We have shown the strong signatures that asymmetric DM particles with low masses and large SD scattering cross sections with baryons produce on low-mass stars. We have focused in the study of the stars KIC 8006161, HD 52265 and $\alpha$ Cen B, revealing large modifications in the central temperatures and densities of the models and the suppression of the convective core expected in 1.1-1.3~M$_{\odot}$ stars.

In the case of $\alpha$ Cen B, we have shown that the asteroseismic parameter $\langle\delta \nu_{02}\rangle$ can be used to impose competitive constraints to the DM characteristics. In particular, we were able to exclude with 95\% confidence level ADM candidates with $m_{\chi}\simeq5\;$GeV and $\sigma_{\chi,SD}\geq3\times10^{-36}\;$cm$^2$. These new constraints are comparable with the present limits from direct detection experiments ($\sigma_{\chi,SD}\gtrsim2\times10^{-37}\;$cm$^2$ for $m_{\chi}\simeq5\;$GeV, see~\cite{2012PhLB..711..153A}) because the sensitivity of the detectors drops at low WIMP masses. 

Interestingly, low-mass WIMPs with similar characteristics have been advocated to explain the signals in the DAMA/LIBRA and CoGeNT detectors in terms of SD collisions. In ADM models the low mass of the WIMPs is strongly motivated because the relic density of DM is determined by the baryon asymmetry of the Universe, leading to $\Omega_{DM}\sim (m_{DM}/m_{b}) \Omega_b$~\citep{Kaplan:2009ag}. Our approach may provide a complementary test of these low-mass WIMP models, in the context of controversy over the incompatible results between different direct detection experiments.

Asteroseismology thus arises as a promising strategy for indirect DM searches. Compared to helioseismology, the asteroseismic searches of DM allow the study of stars with masses lower than that of the Sun, which are more strongly influenced by the additional cooling mechanism provided by the DM particles. In addition, the asteroseismic confirmation of the presence or the absence of convective cores in 1.1-1.3~M$_{\odot}$ stars, such as HD 52265, may provide further constraints on the nature of DM.

The future perspectives of this approach are also exciting. If the small frequency spacings are identified in the oscillations of stars located in environments with high expected DM densities, such as globular clusters, then the sensitivity of the approach proposed in this work will reach much smaller WIMP-proton scattering cross sections and larger WIMP masses. Moreover, in the event of a successful identification of the properties of DM after hypothetical positive results in different experiments, asteroseismology may allow the determination of the density of DM at any specific location where a star is observed.

\acknowledgments
\textit{Acknowledgments}: We are grateful to the authors of DarkSUSY, CESAM and ADIPLS for making their codes publicly available. JC thanks A.~Serenelli and P.~Tinyakov for helpful comments. This work was supported by grants from FCT-MCTES (SFRH/BD/44321/2008) and Funda\c c\~ao Calouste Gulbenkian.


\begin{thebibliography}{50}
\expandafter\ifx\csname natexlab\endcsname\relax\def\natexlab#1{#1}\fi

\bibitem[{{Archambault} {et~al.}(2012){Archambault}, {Behnke}, {Bhattacharjee},
  {et~al.}}]{2012PhLB..711..153A}
{Archambault}, S., {Behnke}, E., {Bhattacharjee}, P., {et~al.} 2012, PhLB, 711,
  153

\bibitem[{Ballot {et~al.}(2011)Ballot, Gizon, Samadi, Vauclair, Benomar,
  {et~al.}}]{Ballot:2011kn}
Ballot, J., Gizon, L., Samadi, R., Vauclair, G., Benomar, O., {et~al.} 2011,
  A\&A, 530, A97

\bibitem[{Bertone(2010)}]{Bertone:2010at}
Bertone, G. 2010, Natur, 468, 389

\bibitem[{Bertone \& Fairbairn(2008)}]{Bertone:2007ae}
Bertone, G. \& Fairbairn, M. 2008, Phys.Rev., D77, 043515

\bibitem[{{Blennow} {et~al.}(2012){Blennow}, {Fernandez Martinez}, {Mena},
  {Redondo}, \& {Serra}}]{Blennow:2012de}
{Blennow}, M., {Fernandez Martinez}, E., {Mena}, O., {Redondo}, J., \& {Serra},
  e.~P. 2012, JCAP, 7, 22

\bibitem[{{Bonaca} {et~al.}(2012){Bonaca}, {Tanner}, {Basu},
  {et~al.}}]{Bonaca:2012av}
{Bonaca}, A., {Tanner}, J.~D., {Basu}, S., {et~al.} 2012, \apjl, 755, L12

\bibitem[{{Bruntt} {et~al.}(2012){Bruntt}, {Basu}, {Smalley},
  {et~al.}}]{Bruntt:2012hs}
{Bruntt}, H., {Basu}, S., {Smalley}, B., {et~al.} 2012, \mnras, 423, 122

\bibitem[{Casanellas \& Lopes(2009)}]{Casanellas:2009dp}
Casanellas, J. \& Lopes, I. 2009, Astrophys.J., 705, 135

\bibitem[{Casanellas \& Lopes(2011{\natexlab{a}})}]{Casanellas:2011qh}
---. 2011{\natexlab{a}}, Astrophys.J., 733, L51

\bibitem[{Casanellas \& Lopes(2011{\natexlab{b}})}]{Casanellas:2010he}
---. 2011{\natexlab{b}}, Mon.Not.Roy.Astron.Soc., 410, 535

\bibitem[{Chaplin {et~al.}(2011)Chaplin, Kjeldsen, Christensen-Dalsgaard,
  {et~al.}}]{Chaplin:2011wa}
Chaplin, W., Kjeldsen, H., Christensen-Dalsgaard, J., {et~al.} 2011, Science,
  332, 213

\bibitem[{{Christensen-Dalsgaard}(2008)}]{art-Ch-Dals2008Ap&SS}
{Christensen-Dalsgaard}, J. 2008, Astrophysics and Space Science, 316, 113

\bibitem[{{C{\'o}rsico} {et~al.}(2012){C{\'o}rsico}, {Althaus}, {Miller
  Bertolami}, {et~al.}}]{Corsico:2012ki}
{C{\'o}rsico}, A.~H., {Althaus}, L.~G., {Miller Bertolami}, M.~M., {et~al.}
  2012, \mnras, 424, 2792

\bibitem[{Cumberbatch {et~al.}(2010)Cumberbatch, Guzik, Silk, Watson, \&
  West}]{Cumberbatch:2010hh}
Cumberbatch, D.~T., Guzik, J., Silk, J., Watson, L.~S., \& West, S.~M. 2010,
  Phys.Rev., D82, 103503

\bibitem[{{Cunha} \& {Brand{\~a}o}(2011)}]{2011A&A...529A..10C}
{Cunha}, M.~S. \& {Brand{\~a}o}, I.~M. 2011, Astronomy and Astrophysics, 529,
  A10

\bibitem[{Davoudiasl {et~al.}(2011)Davoudiasl, Morrissey, Sigurdson, \&
  Tulin}]{Davoudiasl:2011fj}
Davoudiasl, H., Morrissey, D.~E., Sigurdson, K., \& Tulin, S. 2011, Phys.Rev.,
  D84, 096008

\bibitem[{{Escobar} {et~al.}(2012){Escobar}, {Th{\'e}ado}, {Vauclair},
  {et~al.}}]{Escobar:2012qb}
{Escobar}, M.~E., {Th{\'e}ado}, S., {Vauclair}, S., {et~al.} 2012, \aap, 543,
  A96

\bibitem[{{Frandsen} \& {Sarkar}(2010)}]{art-FrandsenSarkar2010PhRvL}
{Frandsen}, M.~T. \& {Sarkar}, S. 2010, Physical Review Letters, 105

\bibitem[{{Garbari} {et~al.}(2012){Garbari}, {Liu}, {Read}, \&
  {Lake}}]{Garbari:2012ff}
{Garbari}, S., {Liu}, C., {Read}, J.~I., \& {Lake}, G. 2012, \mnras, 425, 1445

\bibitem[{Garcia {et~al.}(2010)Garcia, Mathur, Salabert,
  {et~al.}}]{Garcia:2010rn}
Garcia, R.~A., Mathur, S., Salabert, D., {et~al.} 2010, Science, 329, 1032

\bibitem[{{Gilliland} {et~al.}(1986){Gilliland}, {Faulkner}, {Press}, \&
  {Spergel}}]{1986ApJ...306..703G}
{Gilliland}, R.~L., {Faulkner}, J., {Press}, W.~H., \& {Spergel}, D.~N. 1986,
  ApJ, 306, 703

\bibitem[{{Gondolo} {et~al.}(2004){Gondolo}, {Edsj{\"o}}, {Ullio},
  {et~al.}}]{art-GondoloEdsjoDarkSusy2004}
{Gondolo}, P., {Edsj{\"o}}, J., {Ullio}, P., {et~al.} 2004, JCAP, 7, 8

\bibitem[{{Gough}(1986)}]{1986hmps.conf..117G}
{Gough}, D.~O. 1986, in Hydrodynamic and Magnetodynamic Problems in the Sun and
  Stars, ed. Y.~{Osaki}, 117

\bibitem[{{Gould}(1987)}]{art-Gould1987}
{Gould}, A. 1987, ApJ, 321, 571

\bibitem[{Gould \& Raffelt(1990)}]{Gould:1989ez}
Gould, A. \& Raffelt, G. 1990, Astrophys.J., 352, 669

\bibitem[{{Griest} \& {Seckel}(1987)}]{1987NuPhB.283..681G}
{Griest}, K. \& {Seckel}, D. 1987, Nuclear Physics B, 283, 681

\bibitem[{{Ilie} {et~al.}(2012){Ilie}, {Freese}, {Valluri}, {Iliev}, \&
  {Shapiro}}]{2012MNRAS.422.2164I}
{Ilie}, C., {Freese}, K., {Valluri}, M., {Iliev}, I.~T., \& {Shapiro}, P.~R.
  2012, Mon.Not.Roy.Astron.Soc., 422, 2164

\bibitem[{Iminniyaz {et~al.}(2011)Iminniyaz, Drees, \& Chen}]{Iminniyaz:2011yp}
Iminniyaz, H., Drees, M., \& Chen, X. 2011, JCAP, 1107, 003

\bibitem[{Iocco {et~al.}(2012)Iocco, Taoso, Leclercq, \& Meynet}]{Iocco:2012wk}
Iocco, F., Taoso, M., Leclercq, F., \& Meynet, G. 2012, Phys.Rev.Lett., 108,
  061301

\bibitem[{Kaplan {et~al.}(2009)Kaplan, Luty, \& Zurek}]{Kaplan:2009ag}
Kaplan, D.~E., Luty, M.~A., \& Zurek, K.~M. 2009, Phys.Rev., D79, 115016

\bibitem[{Kjeldsen {et~al.}(2005)Kjeldsen, Bedding, Butler,
  {et~al.}}]{Kjeldsen:2005td}
Kjeldsen, H., Bedding, T.~R., Butler, R.~P., {et~al.} 2005, Astrophys.J., 635,
  1281

\bibitem[{Kouvaris \& Tinyakov(2011)}]{Kouvaris:2011fi}
Kouvaris, C. \& Tinyakov, P. 2011, Phys.Rev.Lett., 107, 091301

\bibitem[{{Leung} {et~al.}(2012){Leung}, {Chu}, \& {Lin}}]{art-LeungChuLin2012}
{Leung}, S.-C., {Chu}, M.-C., \& {Lin}, L.-M. 2012, \prd, 85, 103528

\bibitem[{Li {et~al.}(2012)Li, Wang, \& Cheng}]{Li:2012qf}
Li, X., Wang, F., \& Cheng, K. 2012, JCAP, 1210, 031

\bibitem[{Lopes {et~al.}(2011)Lopes, Casanellas, \& Eugenio}]{Lopes:2011rx}
Lopes, I., Casanellas, J., \& Eugenio, D. 2011, Phys.Rev., D83, 063521

\bibitem[{Lopes \& Silk(2010)}]{Lopes:2010fx}
Lopes, I. \& Silk, J. 2010, Astrophys.J., 722, L95

\bibitem[{Lopes \& Silk(2012)}]{Lopes:2012af}
---. 2012, Astrophys.J., 757, 130

\bibitem[{Mathur {et~al.}(2012)Mathur, Metcalfe, Woitaszek,
  {et~al.}}]{Mathur:2012sk}
Mathur, S., Metcalfe, T., Woitaszek, M., {et~al.} 2012, Astrophys.J., 749, 152

\bibitem[{Michel {et~al.}(2008)Michel, Baglin, Auvergne, Catala, \&
  Samadi}]{Michel:2008im}
Michel, E., Baglin, A., Auvergne, M., Catala, C., \& Samadi, R. 2008, Science,
  322, 558

\bibitem[{{Morel}(1997)}]{1997A&AS..124..597M}
{Morel}, P. 1997, A\&AS, 124, 597

\bibitem[{{Renzini}(1987)}]{1987A&A...171..121R}
{Renzini}, A. 1987, A\&A, 171, 121

\bibitem[{Scott {et~al.}(2009)Scott, Fairbairn, \& Edsjo}]{Scott:2008ns}
Scott, P., Fairbairn, M., \& Edsjo, J. 2009, Mon.Not.Roy.Astron.Soc., 394, 82

\bibitem[{Scott {et~al.}(2011)Scott, Venkatesan, Roebber,
  {et~al.}}]{Scott:2011ni}
Scott, P., Venkatesan, A., Roebber, E., {et~al.} 2011, Astrophys.J., 742, 129

\bibitem[{{Silva Aguirre} {et~al.}(2011){Silva Aguirre}, {Ballot}, {Serenelli},
  \& {Weiss}}]{2011A&A...529A..63S}
{Silva Aguirre}, V., {Ballot}, J., {Serenelli}, A.~M., \& {Weiss}, A. 2011,
  Astronomy and Astrophysics, 529, A63

\bibitem[{Sivertsson \& Gondolo(2011)}]{Sivertsson:2010zm}
Sivertsson, S. \& Gondolo, P. 2011, Astrophys.J., 729, 51

\bibitem[{Spolyar {et~al.}(2008)Spolyar, Freese, \& Gondolo}]{Spolyar:2007qv}
Spolyar, D., Freese, K., \& Gondolo, P. 2008, Phys.Rev.Lett., 100, 051101

\bibitem[{{Taoso} {et~al.}(2010){Taoso}, {Iocco}, {Meynet}, {Bertone}, \&
  {Eggenberger}}]{art-Taosoetal2010PhRvD}
{Taoso}, M., {Iocco}, F., {Meynet}, G., {Bertone}, G., \& {Eggenberger}, P.
  2010, Phys.Rev.D, 82

\bibitem[{{Turck-Chi{\`e}ze} {et~al.}(2012){Turck-Chi{\`e}ze}, {Garc{\'{\i}}a},
  {Lopes}, {Ballot}, {Couvidat}, {Mathur}, {Salabert}, \&
  {Silk}}]{2012ApJ...746L..12T}
{Turck-Chi{\`e}ze}, S., {Garc{\'{\i}}a}, R.~A., {Lopes}, I., {Ballot}, J.,
  {Couvidat}, S., {Mathur}, S., {Salabert}, D., \& {Silk}, J. 2012, Astrophys.
  J., 746, L12

\bibitem[{Zackrisson {et~al.}(2010)Zackrisson, Scott, Rydberg,
  {et~al.}}]{Zackrisson:2010jd}
Zackrisson, E., Scott, P., Rydberg, C.-E., {et~al.} 2010, Astrophys.J., 717,
  257

\bibitem[{Zentner \& Hearin(2011)}]{Zentner:2011wx}
Zentner, A.~R. \& Hearin, A.~P. 2011, Phys.Rev., D84, 101302

\end{thebibliography}

\end{document}